# Mode Error Analysis of Impedance Measurement using Twin Wires *

Huang Liang-Sheng(黄良生)[a,b], Yoshiro Irie(入江吉郎)[a,c], Liu Yu-Dong(刘瑜冬)[a,b],

Wang Sheng(王生)[a,b, #]

[a]China Spallation Neutron Source (CSNS), Institute of High Energy Physics (IHEP), Chinese Academy of Sciences (CAS)，Dongguan 523803, China

[b]Dongguan Institute of Neutron Science (DINS), Dongguan 523808, China

[c]KEK, High Energy Accelerator Research Organization, 1-1 Oho, Tsukuba, Ibaraki 305-0801, Japan

ABSTRACT

Both longitudinal and transverse coupling impedance for some critical components need to be measured for accelerator design. The twin wires method is widely used to measure longitudinal and transverse impedance on the bench. A mode error is induced when the twin wires method is used with a two-port network analyzer. Here, the mode error is analyzed theoretically and an example analysis is given. Moreover, the mode error in the measurement is a few percent when a hybrid with no less than 25 dB isolation and a splitter with no less than 20 dB magnitude error are used.

## 1. Introduction

The electromagnetic interaction of a charged particle beam with its surroundings in the accelerator is conveniently described by the coupling impedances of its components. Coupling impedance is basically an engineering concept, since it is defined as the ratio of voltage divided by current. Coupling impedance may lead to heating of components and limit the increase of beam power, so it should usually be measured on the bench. For transverse coupling impedance measurement with wire methods, a conventional way, the twin wires method, is to insert two parallel wires with out of phase wire current (differential-mode signal) into the device under test (DUT) to produce a dipole current moment. Transverse impedance by the twin wires method was firstly measured by G. Nassibian in 1978 [1]. There has been a long history in the development of this method and in the improvement of its accuracy in both theory and technique. The twin wires method has therefore become a standard method for transverse impedance measurement [2][3][4]. The twin wires method has recently been developed further recently in longitudinal coupling impedance measurement with in phase wire current (common-mode signal) [5][6]. Therefore, it is convenient to use the twin wires method to measure longitudinal and transverse impedance at the same time.

The basic principle of impedance measurement with the wire method is that the fields of an ultrarelativistic beam on the beam tube wall can be simulated by the propagation of a time-harmonic TEM mode in the transmission line. In the typical bench measurement involving a network analyzer, the impedance measurements are made using the twin wires method to obtain the forward scattering coefficient $S_{DUT,21}$ of the DUT with respect to a reference (REF) tube of equal length and interpreted via

※Supported by National Natural Science Foundation of China (11175193, 11275221)

#wangs@ihep.ac.cn

the log-formula [3]

$$Z_L = -2Z_c^{cm} \ln(\frac{S_{DUT,21}}{S_{REF,21}}), \tag{1}$$

$$Z_T = -2Z_c^{dm} \frac{c}{\omega(2d)^2} \ln(\frac{S_{DUT,21}}{S_{REF,21}}), \tag{2}$$

With the speed of light $c$, frequency $\omega$, the spacing of the twin wires $2d$, and the characteristic impedance of a differential-mode line [7]

$$Z_c^{dm} = \frac{1}{\pi}\sqrt{\frac{\mu_0}{\varepsilon_0}} \ln\left(\frac{d+\sqrt{d^2-a^2}}{a} \frac{b^2-d\sqrt{d^2-a^2}}{b^2+d\sqrt{d^2-a^2}}\right), \tag{3}$$

where $\mu_0$ and $\varepsilon_0$ are the permeability and permittivity in free space, $a$ and $b$ are the radius of the wire and beam tube respectively. Based on the same process, the characteristic impedance of common-mode transmission line can be expressed as

$$Z_c^{cm} = \frac{1}{4\pi}\sqrt{\frac{\mu_0}{\varepsilon_0}} \ln\frac{b^2-d^2}{a\cdot(2d-a)}. \tag{4}$$

One considers the primary modes to be common-mode in longitudinal measurement and differential-mode in transverse measurement, and the secondary modes to be differential mode in longitudinal measurement and common mode in transverse measurement. The primary mode is the desired mode and the secondary mode is defined as the mode error. The point of impedance measurement by the twin wires method produces purely common-mode and differential-mode wire current. Some network analyzers with multiple test ports can generate pure primary mode without mode error in the measurement, e. g. HP 8753E and Rohde and Schwarz ZVT-8 [5], the measurement may be restricted, however, in network analyzers with only two test ports, and secondary mode signal is inevitably generated in the production of the primary mode signal when such an analyzer used, producing mode error in the impedance measurement. Common-mode and differential-mode signals are offered by mode generators, such as amplifier circuits, transformers, in phase splitters (0° splitters) and out of phase splitters (180° splitters, which are usually called hybrids). The mode error depends on the mode generator. Therefore, it is necessary to use a high quality generator to produce the desired mode, but the mode error should be analyzed at the same time. Some other studies give more details of the mode generator [8][9][10], so the mode error is analyzed in this paper as the complementarities for longitudinal and transverse impedance measurement by the twin wires method. The generator is typically hybrid and splitter, so the mode errors of the generator, such as isolation, magnitude error, are firstly introduced in Section 2. Then, the mode error of the forward scattering coefficient is analyzed theoretically in the impedance measurement in Section 3. Based on a typical hybrid and splitter, an example of mode error in the measurement is given in Section 4; according to the example, a generator with no more than 25 dB mode errors should be used in the measurement if the mode-error of the measured impedance is to be restricted to a few percent.

## 2. Mode Error of the Generator

The mode error of the measured impedance relates to the mode error of the mode generator, which depends on two output $S_{21}$ magnitudes. To confirm the mode error of the impedance, the mode error of the generator is first measured.

The generator of the common-mode and differential-mode signals is shown in Figure 1. The input signal at the input port ('S') is $U_i$, and the output signals at the output ports ('1' and '2') are $U_1$ and $U_2$.

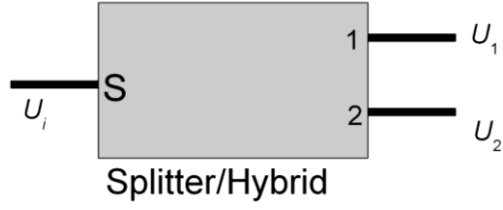

Figure 1: The generation of common-mode or differential-mode signals

The hybrid produces a differential-mode signal, and its mode error is a common-mode signal. The mode error ($\varepsilon_{21}^{cc}$) of the hybrid is generally isolation [11], which is the forward scattering coefficient S$_{21}$ from one output port of the hybrid to the other when the *port S* matches.

The common-mode signal is produced by the splitter, and its mode error is a differential mode signal. The splitter is different from the hybrid. One output port of the splitter is usually not isolated from the other, so the mode error is considered as a magnitude error. The splitter output S$_{21}$ magnitudes have two parts in Figure 1: S$_{21}$ magnitude from *port S* to *port 1* ($S_{21}^{i1}$), and from *port S* to *port 2* ($S_{21}^{i2}$). $S_{21}^{i1}$ is measured by the network analyzer, and *port S* and *port 1* connect with TEST PORT 1 and TEST PORT 2 of the analyzer respectively when *port 2* matches. $S_{21}^{i2}$ is also measured when *port S*, *port 2* and *port 1* connect with TEST PORT 1, TEST PORT 2 and LOAD respectively. The measured S$_{21}$ consists of desired mode S$_{21}$ and the mode error, and is expressed as

$$S_{21}^{i1} = S_{21}^{cm} + \varepsilon_{21}^{dc},$$ (5)

$$S_{21}^{i2} = S_{21}^{cm} - \varepsilon_{21}^{dc},$$ (6)

where, the magnitude error of the splitter is $\varepsilon_{21}^{dc}$, which can be obtained from Eq. (5) and Eq. (6) as

$$\varepsilon_{21}^{dc} = \left| S_{21}^{i1} - S_{21}^{i2} \right| / 2.$$ (7)

### 3. Analysis of S₂₁ Mode Error

Unwanted modes in the network analyzer with two test ports cause mode error when the impedance is measured with the twin wires method, so the mode error should be given in the measurement. The mode error is analyzed as below based on the splitter and the hybrid. The schematic for the impedance measurement is shown in Figure 2. The network analyzer with two test ports is used, the generator (hybrid/splitter) on the left being used to produce the common-mode or the differential-mode wire current, which is then combined by the combiner (hybrid/splitter) on the right. The calibration plane for the measurement is chosen as the output plane of the generator.

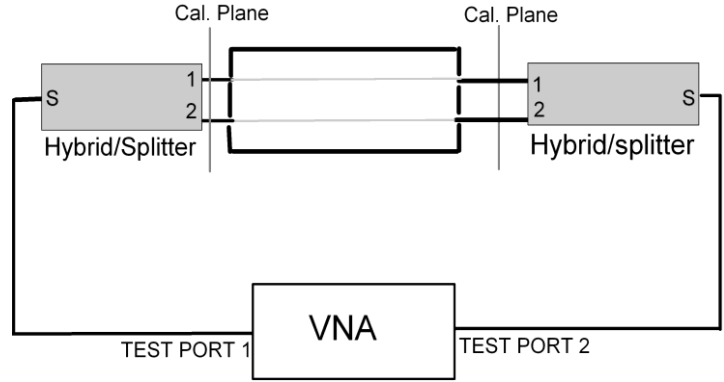

Figure 2: The schematic of mode error analysis for two-port network analyzer

We define $S_{21}^{dd}$ as the measured differential-mode forward scattering coefficients when the hybrid is used to produce a differential-mode signal, and $S_{21}^{cc}$ as the measured common-mode scattering coefficients when the splitter is used. Furthermore, these can be expressed as

$$S_{21}^{dd} = S_{dd21} \pm 2S_{cc21}\left(\varepsilon_{21}^{cc}\right)^2,$$
$$S_{21}^{cc} = S_{cc21} \pm 2S_{dd21}\left(\varepsilon_{21}^{dd}\right)^2. \tag{8}$$

In Eq. (8), $S_{dd21}$ is the pure differential-mode scattering coefficient in the measurement, and $S_{cc21}$ is the pure common-mode scattering coefficient. One assumption of Eq. (8) is that the mode error of the generator on the left is the same as that of the combiner on the right. The second terms in the right-hand of the two formulas in Eq. (8) are the mode errors of the measured scattering coefficients. The measured impedance with the mode error is generally obtained by substituting $S_{21}^{cc}$ and $S_{21}^{dd}$ into Eq. (1) and Eq. (2), and the improved impedance without mode error is given by replacing $S_{21}^{cc}$ and $S_{21}^{dd}$ by $S_{cc21}$ and $S_{dd21}$, the different of two impedances is the mode error of the measured impedance, and the relative mode error of impedance can be expressed, based on Eq. (1) and Eq. (2), as [12]

$$\Delta = \frac{2}{S_{21} \cdot \ln(S_{21})} \left| \frac{\Delta S_{21}}{S_{21}} \right|. \tag{9}$$

Here, $S_{21}$ is $S_{21}^{dd}$ and $S_{21}^{cc}$ for the longitudinal and transverse measurements respectively.

## 4. Example Analysis

An example of the mode error analysis is given in the section, using a splitter (ZFRSC-42-S+) and a hybrid (ZFSCJ-2-1) from Mini-circuits [13]. The mode errors of the splitter and the hybrid are first measured, and then the corrected scattering coefficient is obtained for a typical component. The mode error of the impedance is also estimated; the mode error of the impedance with the differential generator mode error are given, which gives a reference for impedance measurement.

The forward scattering coefficients of the hybrid are measured by the network analyzer as shown in the left-hand side of Figure 3. The $S_{21}$ magnitude is not accurate when the frequency is zero, as it is limited by the start frequency of the analyzer. It is clearly shown that the difference in the magnitude is small. The linear isolation is given in the right-hand part of Figure 3, and is bigger than 28 dB. The $S_{21}$ magnitudes ($S_{21}^{i1}$, $S_{21}^{i2}$) of the two output ports for the splitter are also measured and shown in the left-hand part of Figure 4. The right-hand plot gives the magnitude error of the splitter, which is slightly higher 50 dB. The magnitude error of the splitter is therefore extremely small.

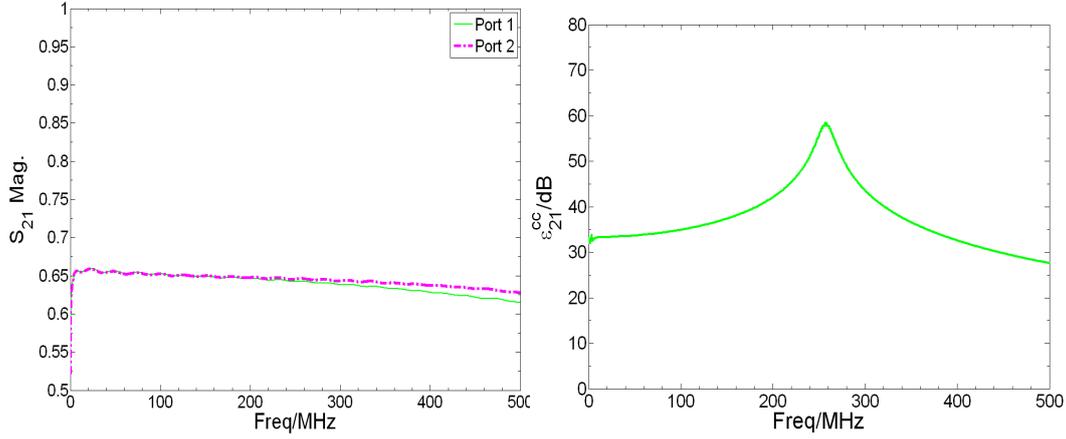

Figure 3: The hybrid $S_{21}$ magnitudes of output ports(left) and isolation (right)

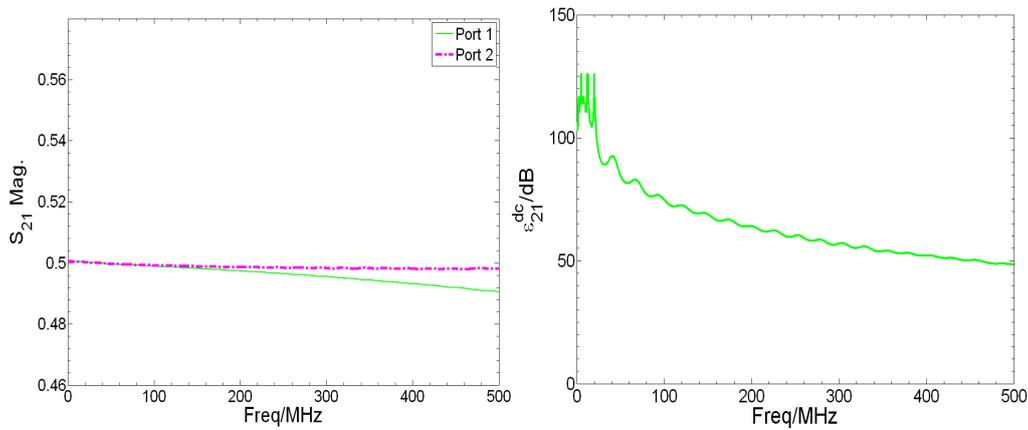

Figure 4: The splitter $S_{21}$ magnitudes (left) and magnitude error (right)

Consider a single localized impedance source - a typical beam pipe, with radius 250 mm and length one meter. 0.25 mm radius copper wires are inserted into the beam pipe and the spacing of the twin wires 40 mm. The characteristic impedances for the common-mode ($Z_c^{cm}$) and differential-mode ($Z_c^{dm}$) given in Eq. (4) and Eq. (3) are 220 $\Omega$ and 603 $\Omega$ respectively. The system impedance of the network analyzer is

$Z_0$=50 Ω. The scattering coefficient $S_{21}$ can be expressed in the microwave transmission line [2] as

$$S_{21} = \frac{2Z_0}{2Z_0 + Z},$$ (10)

where, $Z$ is the characteristic impedance $Z_c^{cm}$ for the longitudinal measurement and $Z_c^{cm}$ for the transverse measurement respectively. Actually, matching is necessary in the impedance measurement, but matching is not discussed here, so the measured system is considered as perfectly matched.

According to Eq. (10), the $S_{21}$ magnitudes for the common-mode and the differential-mode are 0.312 and 0.142, respectively. Substituting the forward scattering coefficients into Eq. (8), the scattering values with mode error are easily obtained. For the longitudinal impedance measurement by the twin wires method, Figure 5 compares the forward scattering coefficient without the mode error and the coefficients with differential mode errors. The deviation of the common-mode scattering coefficient for the splitter (ZFRSC-42-S+) used is very small, being about 3 % for this 14 dB splitter. The differential-mode scattering coefficients with differential hybrid isolations are presented in Figure 6. The error of the scattering coefficient for the hybrid (ZFSCJ-2-1) is about 0.7 %, but goes up to 7 % when 21 dB hybrid isolations is used in the measurement.

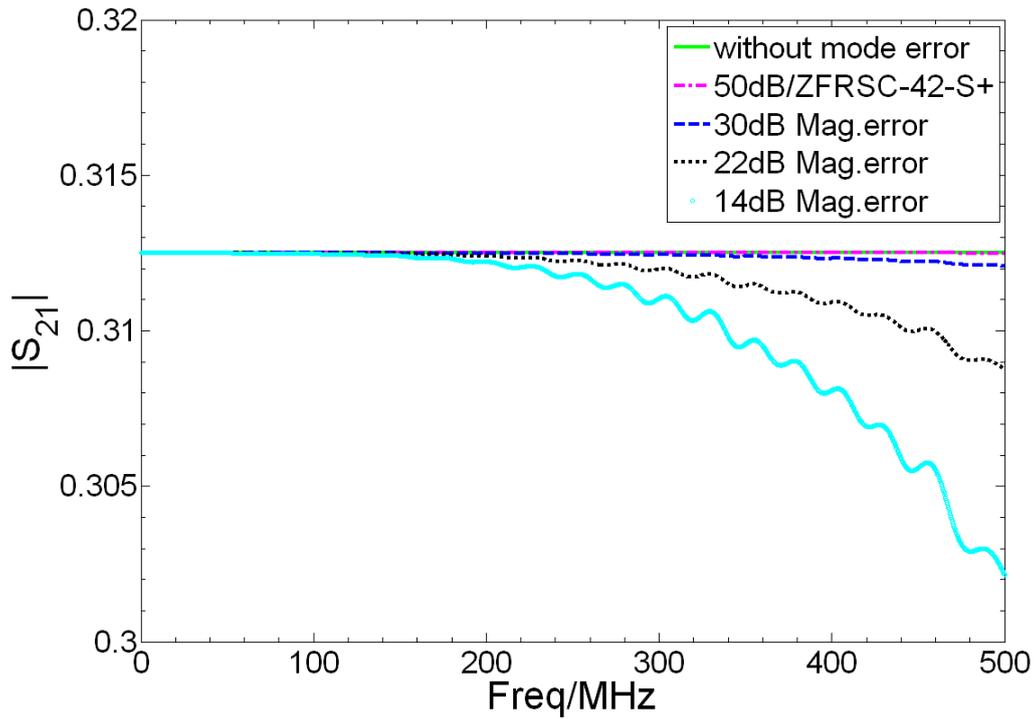

Figure 5: Raw data ($S_{21}$magnitude) without mode error of longitudinal impedance measurement, compared with differential magnitude errors of the splitter

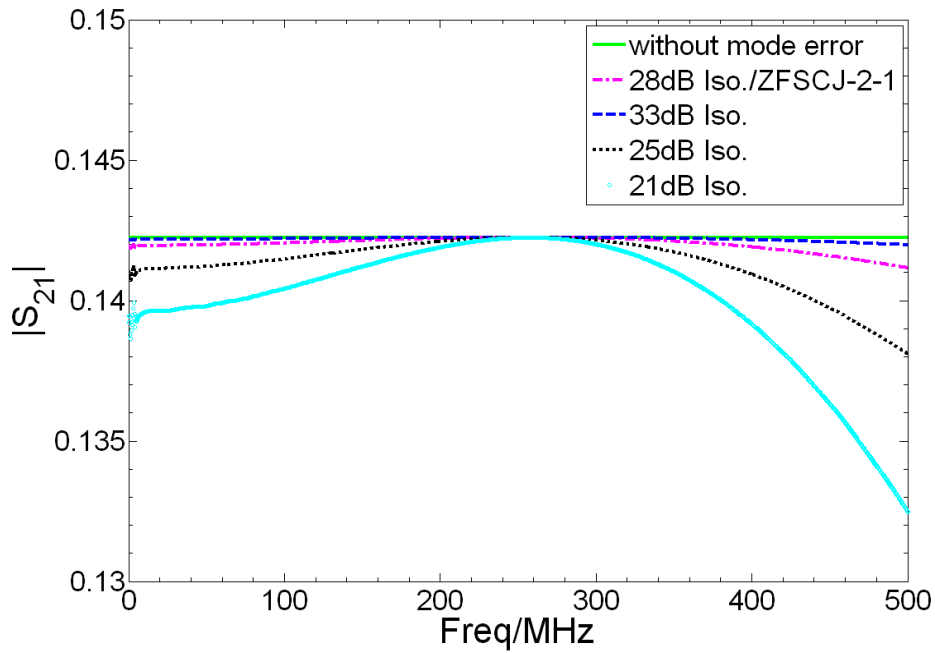

Figure 6: Raw data ($S_{21}$magnitude) without mode error of transverse impedance measurement, compared with differential hybrid mode errors

The mode error of the measured longitudinal and transverse impedance can be estimated through Eq. (9). The relative differential-mode error of the longitudinal impedance is shown in Figure 7. The different mode error in the longitudinal impedance produces different splitter magnitude errors, and the magnitude error of the splitters should be bigger than 20 dB if the error is limited to few percent. The mode error of the splitter (ZFRSC-42-S+) is bigger than 50 dB, and the mode error of impedance is much smaller than 0.1 %, so the error is ignored in this example. The relative common-mode error of transverse impedance with different hybrid isolation is shown in Figure 8, showing that the isolation of the hybrid should be bigger than 25 dB if the error of the transverse impedance is a few percent. The hybrid with 28 dB isolation is used in the example, and the mode error of transverse impedance is about 5 %.

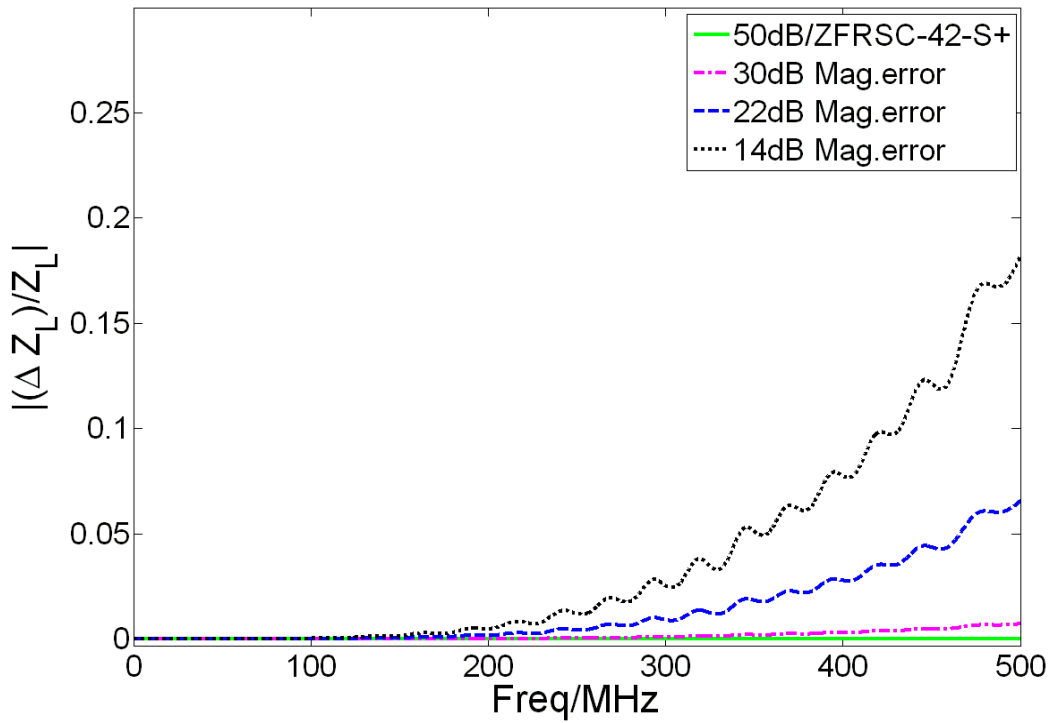

Figure 7: Relative mode error of longitudinal impedance with different splitter magnitude errors

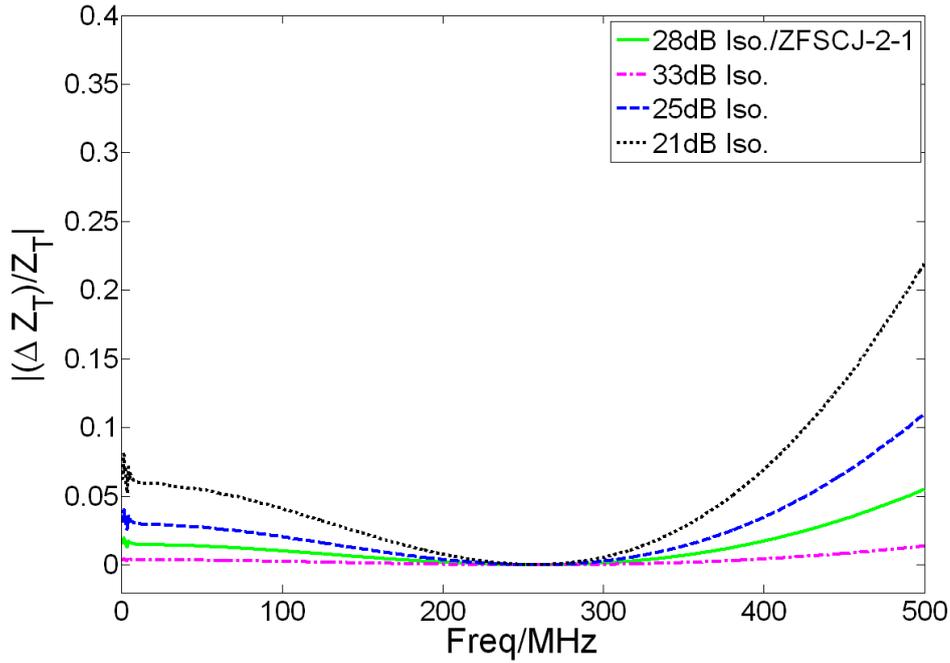

Figure 8: Relative mode error of transverse impedance with different hybrid isolations

## 5. Conclusion

The longitudinal and transverse impedance can be measured by the twin wire method at the same time. The mode error of measured impedance by the network analyzer with two ports is discussed. To restrict the mode error to a few percent, a hybrid with no less than 25 dB isolation and a splitter with no less than 20 dB magnitude error should be used.

**Acknowledgement**

We would like to acknowledge the discussion with Dr. Takeshi Toyama for the measurement of the twin wires method.

References

[1]   G. Nassibian and F. Sacherer, Methods for Measurement Transverse Coupling Impedance in Circular Accelerator, Nucl. Instr. and Methods. 159 (1979): 21-27.

[2]   F. Caspers, BENCH METHODS FOR BEAM-COUPLING IMPEDANCE MEASUREMENT, CERN PS/88-59, Geneva, 1988.

[3]   L.S. Walling, et. al., Transmission line impedance measurements for an advanced hadron facility, Nucl. Instr. and Meth. A 281 (1989), pp. 433.

[4]   G. Arduini, et. al., MEASUREMENTS OF THE SPS TRANSVERSE IMPEDANCE IN 2000, Proceedings of the PAC 2001, Chicago, pp. 2054.

[5]   Takeshi Toyama, Yoshinori Hashimoto, Yoshihisa Shirakabe, COUPLING IMPEDANCE OF THE J-PARC KICKER MAGNETS, Proceedings of HB2006, Tsukuba, Japan, pp. 140.

[6]   Takeshi Toyama, Coupling impedance bench measurement, private connection, 2013.08.20.

[7]   A.W. Gent, Electrical Comm. 33 (1956), 234.

[8]   J. G. Wang, S. Y. Zhang, Coupling impedance measurements of a model fast extraction kicker magnet for the SNS accumulator ring, Nucl. Instr. & Phys. Res., Sect. A 522 (2004): 178-189.

[9]   H. Hahn and D. Davino, Transverse Coupling Impedance of the RHIC Abort Kicker, C-A/AP/52, 2001.05.

[10]  H. Hahn, Impedance measurements of the Spallation Neutron Source extraction kicker system, PHYSICAL REVIEW SPECIAL TOPICS – A & B, VOL. 7, 103501 (2004).

[11]  A. Chao and M. Tigner, Handbook of Accelerator Physics and Engineering, World Scientific, Singapore (1998), pp. 576.

[12]  H. H. Ku, Notes on the use of propagation of error formulas, JOURNAL OF RESEARCH of the National Bureau of Standards - C. Engineering and Instrumentation, Vol. 70C, No.4, 1966.10, pp. 263.

[13]  http://www.minicircuits.com/homepage/homepage.html, 2014.03.10.